\pdfoutput=1

\documentclass[11pt]{article}

\usepackage[]{acl}

\usepackage{times}
\usepackage{latexsym}
\usepackage{amsmath}
\usepackage{amsfonts}
\usepackage{graphicx}
\usepackage{tabularx}
\usepackage{float}
\usepackage{cleveref}
\usepackage{hyperref}
\usepackage{subfigure}
\usepackage{booktabs} 
\usepackage{algorithm,algpseudocode}
\usepackage{comment}
\usepackage{tikz}
\usetikzlibrary{bayesnet}

\usepackage[T1]{fontenc}

\usepackage[utf8]{inputenc}

\usepackage{microtype}

\title{A Joint Learning Approach for Semi-supervised Neural Topic Modeling}

\author{Jeffrey Chiu\Thanks{ Equal contribution} \; 
    Rajat Mittal\footnotemark[1] \;
    Neehal Tumma\footnotemark[1] \;
    Abhishek Sharma \;
    Finale Doshi-Velez\\
  Harvard University, Cambridge, MA\\
}

\begin{document}
\maketitle
\begin{abstract}
Topic models are some of the most popular ways to represent textual data in an interpret-able manner. Recently, advances in deep generative models, specifically auto-encoding variational Bayes (AEVB), have led to the introduction of unsupervised neural topic models, which leverage deep generative models as opposed to traditional statistics-based topic models. We extend upon these neural topic models by introducing the \emph{Label-Indexed Neural Topic Model (LI-NTM)}, which is, to the extent of our knowledge, the first effective upstream semi-supervised neural topic model. We find that LI-NTM outperforms existing neural topic models in document reconstruction benchmarks, with the most notable results in low labeled data regimes and for data-sets with informative labels; furthermore, our jointly learned classifier outperforms baseline classifiers in ablation studies.
\end{abstract}

\section{Introduction}

Topic models are one of the most widely used and studied text modeling techniques, both because of their intuitive generative process and interpretable results \cite{blei2012probabilistic}. Though topic models are mostly used on textual data \cite{rosen2012author, yan2013biterm}, use cases have since expanded to areas such as genomics modeling \cite{liu2016overview} and molecular modeling \cite{doi:10.1021/acs.jcim.7b00249}.



Recently, neural topic models, which leverage deep generative models have been used successfully for learning these probabilistic models. A lot of this success is due to the development of variational autoencoders \cite{rezende2014stochastic,kingma2014autoencoding} which allow for inference of intractable distributions over latent variables through a back-propagation over an inference network. Furthermore, recent research shows promising results for Neural Topic Models compared to traditional topic models due to the added expressivity from neural representations; specifically, we see significant improvements in low data regimes \cite{srivastava2017autoencoding, iwata2021few}.

\begin{figure}[t!]
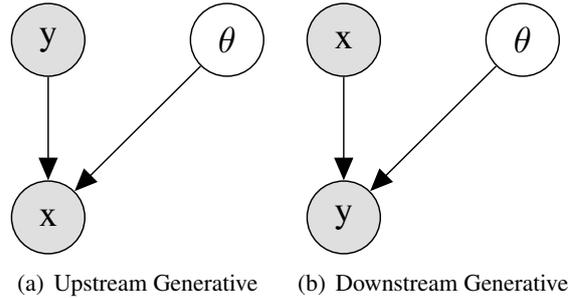

\centering
\subfigure[Upstream Generative]{
 \resizebox{0.225\textwidth}{!}{
      \tikz{ %
        \node[obs] (y) {y} ; %
        \node[latent, right=of y] (z) {$\theta$} ; %
        \node[obs, below=of y] (x) {x} ;
        \edge {y} {x} ; %
        \edge {z} {x} ; %
      }
  }
}
\subfigure[Downstream Generative]{
 \resizebox{0.225\textwidth}{!}{
      \tikz{ %
        \node[obs] (x) {x} ; %
        \node[latent, right=of x] (z) {$\theta$} ; %
        \node[obs, below=of x] (y) {y} ;
        \edge {x} {y} ; %
        \edge {z} {y} ; %
      }
  }
}
\caption{Generative process for downstream vs upstream supervision. Note that in upstream supervision, the label, $y$, supervises the document, $x$, whereas in downstream supervision the document supervises the label. $\theta$ is an arbitrary latent variable, in our case representing topic proportions.}
\label{fig:1}
\end{figure}

Joint learning of topics and other tasks have been researched in the past, specifically through supervised topic models \cite{blei2010supervised, huh2012discriminative, cao2015novel, wang2020neural}. These works are centered around the idea of a prediction task using a topic model as a dimensionality reduction tool. Fundamentally, they follow a downstream task setting (\autoref{fig:1}), where the label is assumed to be generated from the latent variable (topics). On the other hand, an upstream setting would be when the input (document) is generated from a combination of the latent variable (topics) and label, which has the benefit of better directly modeling how the label affects the document, resulting in topic with additional information being injected from the label information. Upstream variants of supervised topic models are much less common, with, to the extent of our knowledge, no neural architectures to this date. \cite{10.5555/1699510.1699543,lacoste2008disclda}.

Our model, the \emph{Label-Indexed Neural Topic Model (LI-NTM)} stands uniquely with respect to all existing topic models. We combine the benefits of an \emph{upstream generative processes} (\autoref{fig:1}), \emph{label-indexed} topics, and a topic model capable of \emph{semi-supervised learning} and \emph{neural topic modeling} to jointly learn a topic model and label classifier. Our main contributions are:
\begin{enumerate}
  \item The introduction of the first upstream semi-supervised neural topic model.
  \item A label-indexed topic model that allows more cohesive and diverse topics by allowing the label of a document to supervise the learned topics in a semi-supervised manner.
  \item A joint training framework that allows for users to tune the trade-off between document classifier and topic quality which results in a classifier that outperforms same classifier trained in an isolated setting for certain hyper-parameters.
\end{enumerate}

\section{Related Work}
\subsection{Neural Topic Models}
Most past work in neural topic models focused on designing inference networks with better model specification in the unsupervised setting. One line of recent research attempts to improve topic model performance by modifying the inference network through changes to the topic priors or regularization over the latent space \cite{miao2016neural, srivastava2017autoencoding, nan2019topic}. Another line of research looks towards incorporating the expressivity of word embeddings to topic models \cite{dieng2019topic, dieng2019dynamic}. 

In contrast to existing work on neural topic models, our approach does not mainly focus on model specification; rather, we create a broader architecture into which neural topic models of all specifications can be trained in an upstream, semi-supervised setting. We believe that our architecture will enable existing neural topic models to be used in a wider range of real-word scenarios where we leverage labeled data alongside unlabeled data and use the knowledge present in document labels to further supervise topic models. Moreover, by directly tying our topic distributions to the labels through label-indexing, we create topics that are specific to labels, making these topics more interpretable as users are directly able to glean what types of documents each of the topics are summarizing.

\subsection{Downstream Supervised Topic Models}
Most supervised topic models follow the downstream supervised framework introduced in s-LDA \cite{blei2010supervised}. This framework assumes a two-stage setting in which a topic model is trained and then a predictive model for the document labels is trained independently of the topic model. Neural topic models following this framework have also been developed, with the predictive model being a discriminative layer attached to the learned topics, essentially treating topic modeling as a dimensionality reduction tool \cite{wang2020neural, cao2015novel, huh2012discriminative}.

In contrast to existing work, \emph{LI-NTM} is an upstream generative model (\autoref{fig:2}, \autoref{fig:3}) following a \emph{prediction-constrained} framework. The upstream setting allows us to implicitly train our classifier and topic model in a \emph{one-stage setting} that is end-to-end. This has the benefit of allowing us to tune the trade-off between our classifier and topic model performance in a \emph{prediction-constrained} framework, which has been shown to achieve better empirical results when latent variable models are used as a dimensionality reduction tool \cite{pmlr-v84-hughes18a, hopeprediction, sharma2021learning}. Furthermore, the upstream setting allows us to introduce the document label classifier as a latent variable, enabling our model to work in semi-supervised settings.

\section{Background}
LI-NTM extends upon two core ideas: Latent Dirichlet Allocation (LDA) and deep generative models. For the rest of the paper, we assume a setting where we have a document corpus of \emph{D} documents, a vocabulary with \emph{V} unique words, and each document having a label from the \emph{L} possible labels. Furthermore let us represent $w_{dn}$ as the $n$-th word in the $d$-th document.
\subsection{Latent Dirichlet Allocation (LDA)} 
LDA is a probabilistic generative model for topic modeling \cite{blei2003latent, blei2010supervised}. Through the process of estimation and inference, LDA learns \emph{K} topics $\beta_{1:K}$. The generative process of LDA posits that each document is a mixture of topics with the topics being global to the entire corpus. For each document, the generative process is listed below:
\begin{enumerate}
  \item Draw topic proportions $\theta_d \sim $ Dirichlet$(\alpha_\theta)$
  \item For each word $w$ in document:
  \begin{enumerate}
     \item Draw topic assignment $z_{dn} \sim $ Cat($\theta_d$)
     \item Draw word $w_{dn} \sim $ Cat($\beta{z_{dn}}$)
  \end{enumerate}
  \item  Draw responses $y|z_{1:N}, \eta, \sigma^2 \sim \mathcal{N}(\eta^T\bar{z}, \sigma^2)$ \emph{(if supervised)}
\end{enumerate}
 where $\bar{z}:= \frac{1}{N}\sum_{i=1}^N z_n$ and the parameters $\eta, \sigma^2$ are estimated during inference. $\alpha_{\theta}$ is a hyper-parameter that serves as a prior for topic mixture proportions. In addition we also have hyper-parameter $\alpha_\beta$ that we use to place a dirichlet prior on our topics, $\beta_k \sim $ Dirichlet$(\alpha_\beta)$.

\subsection{Deep Generative Models}
Deep Generative Models serve as the bridge between probabilistic models and neural networks. Specifically, deep generative models treat the parameters of distributions within probabilistic models as outputs of neural networks. Deep generative models fundamentally work because of the \emph{re-parameterization trick} that allows for backpropogation through Monte-Carlo samples of distributions from the location-scale family. Specifically, for any distribution $g(\cdot)$ from the location-scale family, we have that
\begin{align*}
  z\sim g(\mu, \sigma^2) &\iff z = \mu + \sigma \cdot \epsilon, \epsilon \sim g(0, 1)
\end{align*}
thus allowing differentiation with respect to $\mu, \sigma^2$.

\begin{figure}[t!]
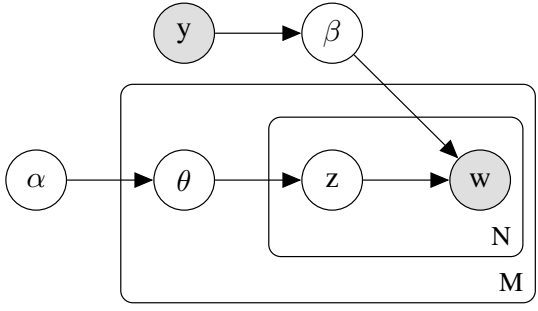

  \centering
  \resizebox{0.45\textwidth}{!}{
      \tikz{ %
        \node[latent] (alpha) {$\alpha$} ; %
        \node[latent, right=of alpha] (theta) {$\theta$} ; %
        \node[latent, right=of theta] (z) {z} ; %
        \node[latent, above=of z] (beta) {$\beta$} ; %
        \node[obs, above=of theta] (y) {y} ; %
        \node[obs, right=of z] (w) {w} ; %
        \plate[inner sep=0.25cm, xshift=-0.12cm, yshift=0.12cm] {plate1} {(z) (w)} {N}; %
        \plate[inner sep=0.25cm, xshift=-0.12cm, yshift=0.12cm] {plate2} {(theta) (plate1)} {M}; %
        \edge {alpha} {theta} ; %
        \edge {theta} {z} ; %
        \edge {z,beta} {w} ; %
        \edge {y} {beta} ; %
      }
  }
  \caption{Generative Process for LI-NTM: The label $y$ indexes into our label-topic-word matrix $\beta$, which is "upstream" of the observed words in the document $w$.}
  \label{fig:2}
\end{figure}
The Variational Auto-encoder is the simplest deep generative model \cite{kingma2014autoencoding} and it's generative process is as follows:
\begin{align*}
    p_\theta(x,z) &= p_\theta(x|z)p(z) \\
    p_\theta(x|z) &\sim \mathcal{N}(\mu_\theta(z), \Sigma_\theta(z)) \\
    p(z) &\sim \mathcal{N}(0, \mathcal{I})
\end{align*}
where $\mu_\theta(z), \Sigma_\theta(z)$ are both parameterized by neural networks with variational parameters $\theta$. Inference on a variational autoencoder is done through approximating the true posterior $p(z|x)$ which is often intractable with an approximation $q_\phi(z|x)$ that is parametrized by a neural network.

The M2 model is the semi-supervised extension of the variational auto-encoder where the input is modeled as being generated by both a continuous latent variable $z$ and the class label $y$ as a latent variable \cite{kingma2014semisupervised}. It follows the generative process below:
\begin{align*}
    p_\theta(x,z,y) &= p_\theta(x|y,z)p(y)p(z) \\
    p_\theta(x|y,z) &\sim \mathcal{N}(\mu_\theta(y,z), \Sigma_\theta(y,z)) \\
    p(y)&\sim Cat(y|\pi) \\
    p(z) &\sim \mathcal{N}(0, \mathcal{I})
\end{align*}
where $\pi$ is parameterizing the distribution on $y$ and $\mu_\theta(y,z), \Sigma_\theta(y,z)$ are both parameterized by neural networks. We then approximate the true posterior $p(y,z | x)$ using by saying 
\begin{align*}
    p(y, z|x) &\approx q_\phi(z|y, x)q_\phi(y|x)
\end{align*}
where $q_\phi(y|x)$ is a classifier that's used in the un-labeled case and $q_\phi(z|y, x)$ is a neural network that takes in the true labels if available and the outputted labels from $q_\phi(y|x)$ if unavailable.

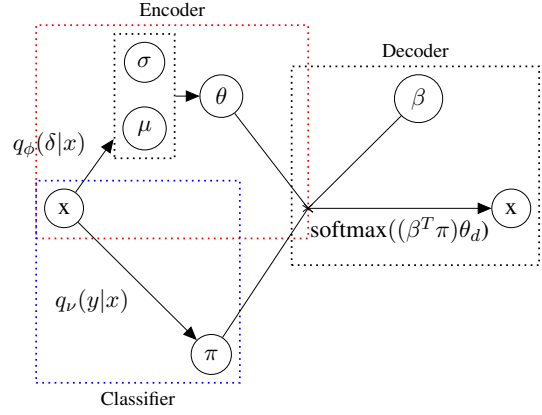
\begin{figure}[t!]
  \centering
      \resizebox{0.45\textwidth}{!}{
          \begin{tikzpicture}
        	\node[circle, draw] (x) {x};
        	\node[circle, draw, right=17em of x] (x2) {x};
        	\node[right=9em of x] (invisible) {};
        	\node[circle, draw, above right =5em of invisible] (beta) {$\beta$};
        	\node[below=1em of invisible] (invisible2){};
        	\node[circle, draw, below right=7em of x] (y) {$\pi$};
        	\node[circle, draw, above right=3em of x] (mu) {$\mu$};
        	\node[circle, draw, above=1em of mu] (sigma) {$\sigma$};
        	\node[draw,dotted,thick,fit=(mu) (sigma)] (dist) {};
        	\node[draw,dotted,thick, red, fit=(dist) (z) (x), label=above:Encoder] (encoder) {};
        	\node[draw,dotted,thick, blue, fit=(y) (x), label=below:Classifier] (classifier) {};
        	\node[draw,dotted, thick, fit=(invisible) (beta) (x2) (invisible2), label=above:Decoder] (decoder) {};
        	\node[circle, draw, right=1em of dist] (z) {$\theta$};
        	\draw[->] (x) -- (y) node[midway, below left] {$q_{\nu}(y|x)$};
        	\draw[->] (x) -- (dist) node[midway, above left] {$q_{\phi}(\delta|x)$};
        	\draw[->] (dist) -- (z);
        	\draw[-|] (z) -- (invisible.center);
        	\draw[-|] (y) -- (invisible.center);
        	\draw[-|] (beta) -- (invisible.center);
        	\draw[->] (invisible.center) -- (x2) node[midway, below] {$\text{softmax}((\beta^T \pi) \theta_d$)};
        \end{tikzpicture}
      }
  \caption{Architecture for LI-NTM in the un-labeled setting. $y$ is used instead of obtaining a probability distribution $\pi$ from the classifier in the labeled setting. $q(\cdot|x)$ are distributions parameterized by neural networks. Note that we can optimize the classifier, encoder, and decoder in one backwards pass.}
  \label{fig:3}
\end{figure}


\section{The Label-Indexed Neural Topic Model}
LI-NTM is a neural topic model that leverages the labels $y$ as a latent variable alongside the topic proportions $\theta$ in generating the document $x$.

Notationally, let us denote the bag of words representation of a document as $x \in \mathbb{R}^V$ and the one-hot encoded document label as $y \in \mathbb{R}^L$. Furthermore, we denote our latent topic proportions as $\theta_d \in \mathbb{R}^K$ and our topics are represented using a three dimensional matrix $\beta \in \mathbb{R}^{L\times K \times V}$.

Under the LI-NTM, the generative process (also depicted in \autoref{fig:2}) of the $d$-th document $x_d$ is the following:
\begin{enumerate}
  \item Draw topic proportions $\theta_d \sim \mathcal{LN}(0, \mathbb{I})$
  \item Draw document label $y_d \sim \pi$
  \item For each word $w$ in document:
  \begin{enumerate}
     \item Draw topic assignment $z_{dn} \sim $ Cat($\theta_d$)
     \item Draw word $w_{dn} \sim $ Cat($\beta_{y_d, z_{dn}}$)
  \end{enumerate}
\end{enumerate}

In Step 1, we draw from the Logistic-Normal $\mathcal{LN(\cdot)}$ to approximate the Dirichlet Distribution while remaining in the location-scale family necessary for re-parameterization \cite{blei2003latent}.  This is done obtained through:
\begin{align*}
    \delta_d &\sim \mathcal{N}(0, \mathbb{I}), \theta_d = softmax(\delta_d)
\end{align*}
Note that since we sample from the Logistic-Normal, we do not require the Dirichlet prior hyper-parameter $\alpha$.

Step 2 is unique for LI-NTM , in the unlabeled case, we sample a label $y_d$ from $\pi$, which is the output of our classifier. In the labeled scenario, we skip step 2 and simply pass in the document label for our $y_d$. Step 3 is typical of traditional LDA, but one key difference is that in step 3b we also index by the $\beta$ by $y_d$ instead of just $z_{dn}$. This step is motivated by how the M2 model extended variational autoencoders to a semi-supervised setting \cite{kingma2014semisupervised}.

A key contribution of our model is the idea of label-indexing. We introduce the supervision of the document labels by having different topics for different labels. Specifically, we have $L \times K$ different topics and we denote the $k$-th topic for label $l$ as the $V$ dimensional vector, $\beta_{l,k}$. Under this setting, we can envision LI-NTM as running a separate LDA for each label once we index our corpus by document labels. 

Label-indexing allows us to effectively train our model in a semi-supervised setting. In the un-labeled data setting, our jointly-learned classifier, $q_\phi(y|x)$, outputs a distribution over the labels, $\pi$. By computing the dot-product between $\pi$ and our topic matrix $\beta$, this allows us to partially index into each label's topic proportional to the classifier's confidence and update the topics based on the unlabeled examples we are currently training on.

\section{Inference and Estimation}
Given a corpus of normalized bag-of-word representation of documents ${x_1, x_2, \cdots, x_d}$ we aim to fit LI-NTM using variational inference in order to approximate intractable posteriors in maximum likelihood estimation \cite{jordan1999introduction}. Furthermore, we amortize the loss to allow for joint learning of the classifier and the topic model.

\subsection{Variational Inference}
We begin first by looking at a family of variational distributions $q_{\phi}(\delta_d| x_d)$ in modeling the untransformed topic proportions and $q_{\nu}(y_d| x_d)$ in modeling the classifier. More specifically, $q_{\phi}(\delta_d| x_d)$ is a Gaussian whose mean and variance are parameterized by neural networks with parameter $\phi$ and $q_{\nu}(y_d| x_d)$ is a distribution over the labels parameterized by a MLP with parameter $\nu$  \cite{kingma2014autoencoding, kingma2014semisupervised}. 

\begin{algorithm}[t]
\caption{Topic Modeling with LI-NTM}
\label{alg:cap}
\begin{algorithmic}
\State Initialize model and variational parameters
\For{iteration $i = 1, 2, \ldots$}
\For{each document $c$ in ${c_1, c_2, \cdots, c_d}$}
\State Get normalized bag-of-word representation\ $x_d$
\State Compute $\mu_d = \textrm{NN}_{encoder}(x_d | \phi_{\mu})$
\State Compute $\Sigma_d = \textrm{NN}_{encoder}(x_d | \phi_{\Sigma})$
\If{labeled} 
    \State $\pi = y_d$
\Else
    \State $\pi = \textrm{NN}_{classifier}(x_d | \nu)$
\EndIf 
\State Sample $\theta_d \sim \mathcal{LN}(\mu_d, \Sigma_d)$
\For{each word in the document}
\State $p(w_{dn} | \theta_d, \pi)=\text{softmax}(\beta)^T \pi \theta_d$
\EndFor
\EndFor
\State Compute the ELBO and its gradient (backprop.)
\State Update model parameters $\beta$
\State Update variational parameters ($\phi_\mu, \phi_\Sigma, \nu$)
\EndFor
\end{algorithmic}
\label{alg:LI-NTM}
\end{algorithm}

We use this family of variational distributions alongside our classifier to lower-bound the marginal likelihood. The evidence lower bound (ELBO) is a function of model and variational parameters and provides a lower bound for the complete data log-likelihood. We derive two ELBO-based loss functions: one for the labeled case and one for the unlabeled case and we compute a linear interpolation of the two for our overall loss function.
\begin{align}
    \mathcal{L}_{u} &= \sum_{d=1}^D \sum_{n=1}^{N_d} \mathbb{E}_q[\log p(w_{dn} | \delta_d, q_{\nu}(y_d|x_d)]\nonumber \\
    &- \tau KL(q_{\phi}(\delta_d| x_d) || p(\delta_d))  \\
    \mathcal{L}_{l} &= \sum_{d=1}^D \sum_{n=1}^{N_d} \mathbb{E}_q[\log p(w_{dn} | \delta_d, q_{\nu}(y_d|x_d)]\nonumber \\
    &- \tau KL(q_{\phi}(\delta_d| x_d) || p(\delta_d)) \nonumber \\
    &+ \rho \mathcal{H}(y_d, q_{\nu}(y_d|x_d))
\end{align}
where Equation 1 serves as our unlabeled loss and Equation 2 serves as our labeled loss. $\mathcal{H}(\cdot, \cdot)$ is the cross-entropy function. $\tau$ and $\rho$ are hyper-parameters on the KL and cross-entropy terms in the loss respectively. 

These hyper-parameters are well motivated. $\tau$ is seen to be a hyper-parameter that tempers our posterior distribution over weights, which has been well-studied and shown to increase robustness to model mis-specification \cite{mandt2016variational, wenzel2020good}. Lower values $\tau$ would result in posterior distributions with higher probability densities around the modes of the posterior. Furthermore, the $\rho$ hyperparameter in our unlabeled loss is the core hyperparameter that makes our model fit the \emph{prediction-constrained} framework, essentially allowing us to trade-off the between classifier and topic modeling performance \cite{pmlr-v84-hughes18a}. Increasing values of $\rho$ corresponds to emphasizing classifier performance over topic modeling performance.

We treat our overall loss as a combination of our labeled and unlabeled loss with $\lambda \in (0,1)$ being a hyper-parameter weighing the labeled and unlabeled loss. $\lambda$ allows us weigh how heavily we want our unlabeled data to influence our models. Example cases where we may want high values of $\lambda$ are when we have poor classifier performance or a disproportionate amount of unlabeled data compared to label data, causing the unlabeled loss to completely outweigh the labeled loss.
\begin{equation} \label{eq:3}
    \mathcal{L} = \lambda \mathcal{L}_l + (1-\lambda) \mathcal{L}_u
\end{equation}

We optimize our loss with respect to both the model and variational parameters and leverage the \emph{reparameterization trick} to perform stochastic optimization \cite{kingma2014autoencoding}. The training procedure is shown in \autoref{alg:LI-NTM} and a visualization of a forward pass is given in \autoref{fig:3}. This loss function allows us to jointly learn our classification and topic modeling elements and we hypothesize that the implicit regularization from joint learning will increase performance for both elements as seen in previous research studies \cite{zweig2013hierarchical}.

\section{Experimental Setup}
We perform an empirical evaluation of LI-NTM with two corpora: a synthetic dataset and \textit{AG News}.

\subsection{Baselines}
We compare our topic model to the Embedded Topic Model (ETM), which is the current state of the art neural topic model that leverages word embeddings alongside variational autoencoders for unsupervised topic modeling  \cite{dieng2019topic}. Further details about ETM are shown in the appendix (\autoref{sssec:etm}). Furthermore, our baseline for our jointly trained classifier is a classifier with the same architecture outside of our jointly trained setting.

\subsection{Synthetic Dataset}

We constructed our synthetic data to evaluate LI-NTM in ideal and worst-case settings. 
\begin{itemize}
    \item \textbf{Ideal Setting}: An ideal setting for LI-NTM consists of a corpus with similar word distributions for documents with the same label and very dissimilar word distributions for documents with different labels
    \item \textbf{Worst Case Setting} worst-case setting for LI-NTM consists of a corpus where the label has little to no correlation with the distribution of words in a document.
\end{itemize}

Since the labels are a fundamental aspect of LI-NTM we wanted to investigate how robust LI-NTM is in a real-word setting, specifically looking at how robust it was to certain types of mis-labeled data points.  By jointly training our classifier with our topic model, we hope that by properly trading off topic quality and classification quality, our model will be more robust to mis-labeled data since we are able to manually tune how much we want to depend on the data labels.

We use the same distributions to generate the documents for both the ideal and worst-case data. In particular, we consider a vocabulary with $V = 20$ words, and a task with $L = 2$ labels. Documents are generated from one of two distributions, $\mathcal{D}_1$ and $\mathcal{D}_2$. $\mathcal{D}_1$ generates documents which have many occurrences of the first 10 words in the vocabulary (and very few occurrences of the last 10 words), while $\mathcal{D}_2$ does the opposite, generating documents which have many occurrences of the last 10 words in the vocabulary (and very few occurrences of the first 10 words). The distributions $\mathcal{D}_1$ and $\mathcal{D}_2$ have parameters which are generated randomly for each trial, although the shape of the distributions is largely the same from trial to trial.

In the ideal case, the label corresponds directly to the distribution from which the document was generated. For the worst-case data, the label is 0 if the number of words in the document is an even number, and 1 otherwise, ensuring there is little to no correlation between label and word distributions in a document. Note that in our synthetic data experiments, all of the data is labeled. The effectiveness of LI-NTM in semi-supervised domains is evaluated in our \textit{AG News} experiments.

\subsection{AG News Dataset}
The \textit{AG News} dataset is a collection of news articles collected from more than 2,000 news sources by ComeToMyHead, an academic news search engine. This dataset includes 118,000 training samples and 7,600 test samples. Each sample is a short text with a single four-class label (one of world, business, sports and science/technology).

\subsection{Evaluation Metrics}

To evaluate our models, we used accuracy as a metric to gauge the quality of the classifier and perplexity to gauge the quality of the model as a whole. We opted to use perplexity as it is a measure for how well the model generalizes to unseen test data.


\section{Synthetic Data Experimental Results}

\begin{table*}
  
  \centering
  \resizebox{\textwidth}{!}{%
  \begin{tabular}{cccccc}
    \toprule
    Total Num. Topics & ETM & Ideal LI-NTM & WC LI-NTM (V1) & WC LI-NTM (V2) & Perplexity Lower Bound \\
    \midrule
    2 & $11.78$ & $11.42$ & $19.71$ & $18.70$ & $-$  \\
    8 & $11.27$ & $10.72$ & $12.83$ & $10.90$ & $-$  \\
    20 & $10.88$ & $10.50$ & $11.20$ & $10.77$ & $9.50$  \\
    \bottomrule
  \end{tabular}
  }
  \caption{Perplexities of LI-NTM (ideal and worst case synthetic data) compared to ETM for a varied number of topics. WC LI-NTM (V1) corresponds to training the model normally in the worst case setting, while WC LI-NTM (V2) corresponds to training with $\rho=0$. Note that LI-NTM is able to outperform ETM in both the ideal and worst case scenarios.} 
  \label{table:perplexity}
 \end{table*}

\begin{table*}
  \centering
  \begin{tabular}{ccccc}
    \toprule
    Total Num. Topics & Worst Case Labels & Ideal Case Labels \\
    \midrule
    2  & $50.2 \pm{0.6}$ & $54.2 \pm{2.0}$   \\
    8 & $50.4 \pm{0.5}$  & $84.3 \pm{8.8}$\\
    20 & $50.4 \pm{0.2}$ & $93.7 \pm{6.2}$\\
    \bottomrule
  \end{tabular}
  \caption{Accuracies of classifier LI-NTM (V2) on ideal case and worst case labels. LI-NTM (V2) is trained only on worst-case labels but evaluated on both worst case and ideal case label test sets. Note that even though $\alpha=0$ and the training set is only worst case labels, the reconstruction loss distantly supervises the classifier to learn the true ideal case labels.}
  \label{table:v2acc}
 \end{table*}

We used our synthetic dataset to examine the performance of LI-NTM relative to ETM in a setting where the label strongly partitions our dataset into subsets that have distinct topics to investigate the effect and robustness of label indexing.

LI-NTM was trained on the fully labeled version of the both the ideal and worse case label synthetic dataset and ETM was trained on the same dataset with the label excluded, as ETM is a unsupervised method. We varied the number of topics in both LI-NTM and ETM to explore realistic settings $K = 2, 8$ and the extreme setting $K = 20$.

\subsection{Effect of Number of Topics}
\textbf{Takeaway: More topics lead to better performance, especially when the label is uninformative.}

First, we note that as we increase the number of topics, the performance of LI-NTM on ideal case labels, LI-NTM on worst case labels, and ETM improves as shown in \autoref{table:perplexity}. This is expected as having more topics gives the model the capacity to learn more diverse topic-word distributions which leads to an improved reconstruction. However, we note that LI-NTM trained on the worst-case labels benefits most from the increase in the number of topics.
\begin{figure}[t!]
  \centering
  \resizebox{0.5\textwidth}{!}{
    \includegraphics{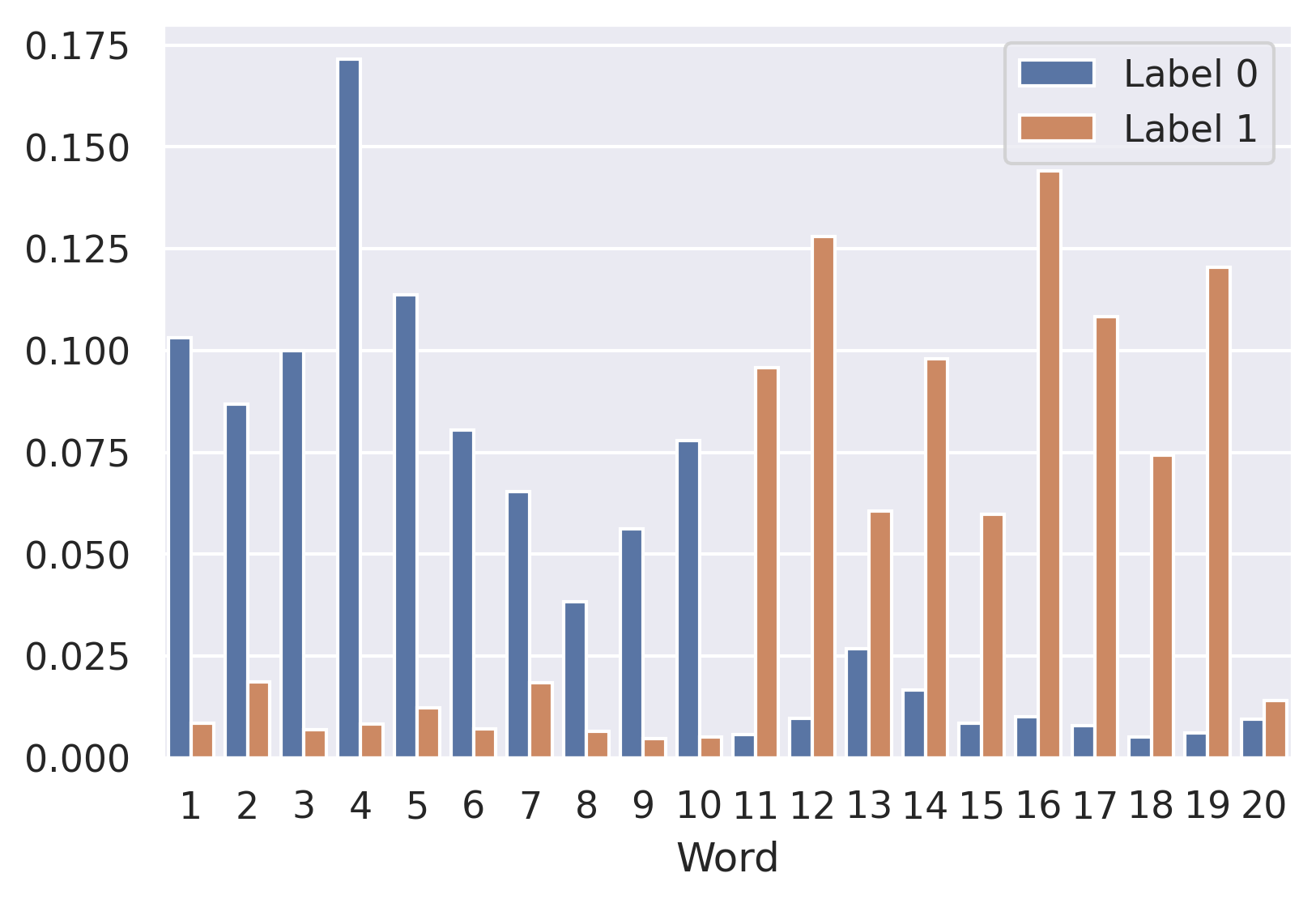}
  }

  \caption{Topic-word probability distribution visualization for LI-NTM on ideal case synthetic dataset with one topic per label. We observe that we learn topics that are strongly label partitioned.}
  \label{fig: labelsynthetic}
\end{figure}

\subsection{Informative Labels}
\textbf{Takeaway: Label Indexing is highly effective when labels partition the dataset well.}
\newline
\newline
\noindent
Next, we note that LI-NTM trained on the ideal case label synthetic dataset outperforms ETM with respect to perplexity (see \autoref{table:perplexity}). This result can be attributed to the fact that LI-NTM leverages label indexing to learn the label-topic-word distribution. Since the ideal case label version of the dataset was constructed such that the label strongly partitions the dataset into two groups (each of which has a very distinct topic-word distribution), and since we had perfect classifier accuracy (the ideal case label dataset was constructed such that the classification problem was trivial), LI-NTM is able to use the output from the classifier to index into the topic-word distribution with 100\% accuracy. 

If we denote the topic-word distribution corresponding to label $0$ by $\beta_0$ and the topic-word distribution corresponding to label $1$ by $\beta_1$, we note that LI-NTM is able to leverage $\beta_0$ to specialize in generating the words for the documents corresponding to label $0$ while using $\beta_1$ to specialize in generating the words for the documents corresponding to label $1$ (see \autoref{fig: labelsynthetic}). Overall, this result suggests that LI-NTM performs well in settings when the dataset exhibits strong label partitioning.

\subsection{Uninformative Labels}
\textbf{Takeaway: With proper hyperparameters, LI-NTM is able to achieve good topic model performance even when we have uninformative labels.}
\newline
\newline
\noindent
We now move to examining the results produced by LI-NTM trained on the worst-case labels. In this data setting, we investigated the robustness of the LI-NTM architecture. Specifically, we looked at a worst-case dataset, where we have labels that are uninformative and are thus not good at partitioning the dataset into subsets that have distinct topics.

In the worst-case setting, we define the following two instances of the LI-NTM model.
\begin{itemize}
  \item \textbf{LI-NTM (V1)} This model refers to the normal ($\rho \neq 0$) version of the model trained in the worst case setting.
  \item \textbf{LI-NTM (V2)} This model refers to a LI-NTM model with zero-ed out classification loss ($\rho = 0$), essentially pushing the model to only accurately reconstruct the original data.
\end{itemize}

\begin{table*}[ht]
  \centering
  \resizebox{\textwidth}{!}{%
  \begin{tabular}{ccccc}
    \toprule
    Data Regime & ETM Perplexity & LI-NTM Perplexity & LI-NTM Accuracy & Baseline Accuracy \\
    \midrule
    5\% labeled, 5\% unlabeled & $\mathbf{205.93}$ & $210.76 \pm 2.17$ & $86.3$\% & $86.2$\%  \\
    5\% labeled, 15\% unlabeled & $190.10$ & $\mathbf{187.66 \pm 2.23}$ & $86.3$\% & $86.2$\%  \\
    5\% labeled, 55\% unlabeled & $177.71$& $\mathbf{175.43 \pm 5.01}$ & $86.8$\% & $86.2$\%  \\
    5\% labeled, 95\% unlabeled & $177.34$& $\mathbf{169.40 \pm 4.08}$ & $87.2$\% & $86.2$\%  \\
    \bottomrule
  \end{tabular}
  }
  \caption{The results from ETM, LI-NTM, and a baseline classifier for the AG News dataset. The baseline classifier was the same for each data regime, hence the duplicate values. Note that in the high data settings, LI-NTM outperformed ETM in terms of perplexity, although in the lowest data setting, the lack of data hurt LI-NTM since it further partitions the topics by labels. Accuracy increased near linearly as unlabeled data increased.}
  \label{agnewsresults}
 \end{table*}
    
 \begin{table*}
    \centering
      \begin{tabular}{cccc}
        \toprule
        Sports & Science/Technology & World & Business \\
        \midrule
        series & web & minister & stocks \\
        game & search & prime & oil \\
        red & google & palestinian & prices \\
        boston & new & gaza & reuters \\
        run & online & israel & company \\
        night & site & leader & shares \\
        league & internet & arafat & inc \\
        yankees & engine & said & percent \\
        new & com & yasser & yesterday \\
        york & yahoo & sharon & percent\\
        \bottomrule
      \end{tabular}
      \caption{Example topics (top ten words) corresponding to each label from LI-NTM run on the AG-News Dataset. Each topic is assigned a label and it is clear that the distribution of words for each topic depends on the label.}
      \label{agnewstopics}
 \end{table*}

For LI-NTM (V1), we did see decreases in performance; namely, that V1 has a worse perplexity than both ETM and ideal case LI-NTM. This aligns with our expectation that having a label with very low correlation to the topic-word distributions in the document results in poor performance in LI-NTM. This can be attributed to the failure of LI-NTM to adequately label-index in cases where this occurs.

However, for LI-NTM (V2) we found that we were actually able to achieve lower perplexity than ETM when the model was told to produce more than 2 topics, even with uninformative labels. To understand why this was happening, we analyzed the accuracy of the original classifier in LI-NTM (V2) on both the worst-case labels (which it was trained on) and the ideal-case labels (which it was not trained on). We report our results in \autoref{table:v2acc}. The key takeaway is that we observed a much higher accuracy on the ideal labels compared to the worst-case labels. This suggests  that when $\rho=0$ the classifier \emph{implicitly} learns the ideal labels that are necessary to learn a good reconstruction of the data, even when the provided labels are heavily \emph{uninformative or misspecified}. This shows the benefit of label-indexing and of jointly learning our topic model and classifier in a semi-supervised fashion. Even in cases with uninformative data points, by setting $\rho = 0$, the joint learning setting of our classifier and topic model pushes the classifier, through the need for successful document reconstruction, to generate a probability distribution over labels that is close to the true, ideal-case labels despite only being given uninformative or mis-labeled data.
 
\section{AG News Experimental Results}

We used the AG News dataset to evaluate the performance of LI-NTM in the semi-supervised setting. Specifically, we aimed to analyze the extent to which unlabeled data can improve the performance of \emph{both} the classifier and topic model in the LI-NTM architecture. Ideally, in the unlabeled case, the distant supervision provided to the classifier from the reconstruction loss would align with the task of predicting correct labels.

We ran four experiments on ETM and LI-NTM in which the amount of unlabeled data was gradually increased, while the amount of labeled data was kept fixed. In each of the experiments, 5\% of the dataset was considered labeled, while 5\%, 15\%, 55\%, and 95\% of the whole dataset was considered unlabeled in each of the four experiments respectively. 
\subsection{Semi-Supervised Learning: Topic Model Performance}
\textbf{Takeaway: Combining label-indexing with semi-supervised learning increases topic model performance.}
\newline
\newline
\noindent
In \autoref{agnewsresults} we observe that perplexity decreases as the model sees more unlabeled data. We also note that LI-NTM has a lower perplexity than ETM in higher data settings, supporting the hypothesis that guiding the reconstruction of a document exclusively via label-specific topics makes reconstruction an easier task. In the lowest data regime (5\% labeled, 5\% unlabeled), LI-NTM performs worse than ETM. This suggests that while in high-data settings, LI-NTM is able to effectively leverage $L = 4$ sets of topics, in low-data settings there are not enough documents to learn sufficient structure. 

\subsection{Semi-Supervised Learning: Classifier Performance}
\textbf{Takeaway: Topic modeling supervises the classifier, resulting in better classification performance.}
\newline
\newline
\noindent
Jointly learning the classifier and topic model also seem to benefit the classifier; \autoref{agnewsresults} shows classification performance increases linearly with the amount of unlabeled data. The accuracy increase suggest the task of reconstructing the bag of words is helpful in news article classification.

Select topics learned from LI-NTM on the AG News Dataset are presented in \autoref{agnewstopics} and the distributions are visualized in the appendix \autoref{fig:topwords}.

\section{Conclusion}
In this paper, we introduced the LI-NTM, which, to the extent of our knowledge, is the first upstream neural topic model with applications to a semi-supervised data setting. Our results show that when applied to both a synthetic dataset and AG News, LI-NTM outperforms ETM with respect to perplexity. Furthermore, we found that the classifier in LI-NTM was able to outperform a baseline that doesn't leverage any unlabeled data. Even more promising is the fact that the classifier in LI-NTM continued to experience gains in accuracy when increasing the proportion of unlabeled data. While we aim to iterate upon our results, our current findings indicate that LI-NTM is comparable with current state-of-the-art models while being applicable in a wider range of real-world settings.

In future work, we hope to further experiment with the idea of label-indexing. While in LI-NTM every topic is label-specific, real datasets have some common words and topics that are label-agnostic. Future work could augment the existing LI-NTM framework with additional label-agnostic global topics which prevent identical topics from being learned across multiple labels. We are also interested in extending our semi-supervised, upstream paradigm to a semi-parametric setting in which the number of topics we learn is not a predefined hyperparameter but rather something that is learned.

\section{Acknowledgements}
AS is supported by R01MH123804, and FDV is supported by NSF IIS-1750358. All authors acknowledge insightful feedback from members of CS282 Fall 2021.

\nocite{*}
\bibliography{acl}

\begin{thebibliography}{36}
\expandafter\ifx\csname natexlab\endcsname\relax\def\natexlab#1{#1}\fi

\bibitem[{Blei(2012)}]{blei2012probabilistic}
David~M Blei. 2012.
\newblock Probabilistic topic models.
\newblock \emph{Communications of the ACM}, 55(4):77--84.

\bibitem[{Blei et~al.()Blei, Griffiths, Jordan, Tenenbaum
  et~al.}]{blei2003hierarchical}
David~M Blei, Thomas~L Griffiths, Michael~I Jordan, Joshua~B Tenenbaum, et~al.
\newblock Hierarchical topic models and the nested chinese restaurant process.

\bibitem[{Blei and Lafferty(2007)}]{blei2007correlated}
David~M Blei and John~D Lafferty. 2007.
\newblock A correlated topic model of science.
\newblock \emph{The annals of applied statistics}, 1(1):17--35.

\bibitem[{Blei and McAuliffe(2010)}]{blei2010supervised}
David~M. Blei and Jon~D. McAuliffe. 2010.
\newblock \href {http://arxiv.org/abs/1003.0783} {Supervised topic models}.

\bibitem[{Blei et~al.(2003)Blei, Ng, and Jordan}]{blei2003latent}
David~M Blei, Andrew~Y Ng, and Michael~I Jordan. 2003.
\newblock Latent dirichlet allocation.
\newblock \emph{the Journal of machine Learning research}, 3:993--1022.

\bibitem[{Cao et~al.(2015)Cao, Li, Liu, Li, and Ji}]{cao2015novel}
Ziqiang Cao, Sujian Li, Yang Liu, Wenjie Li, and Heng Ji. 2015.
\newblock A novel neural topic model and its supervised extension.
\newblock In \emph{Proceedings of the AAAI Conference on Artificial
  Intelligence}, volume~29.

\bibitem[{Dieng et~al.(2019{\natexlab{a}})Dieng, Ruiz, and
  Blei}]{dieng2019topic}
Adji~B. Dieng, Francisco J.~R. Ruiz, and David~M. Blei. 2019{\natexlab{a}}.
\newblock \href {http://arxiv.org/abs/1907.04907} {Topic modeling in embedding
  spaces}.

\bibitem[{Dieng et~al.(2019{\natexlab{b}})Dieng, Ruiz, and
  Blei}]{dieng2019dynamic}
Adji~B Dieng, Francisco~JR Ruiz, and David~M Blei. 2019{\natexlab{b}}.
\newblock The dynamic embedded topic model.
\newblock \emph{arXiv preprint arXiv:1907.05545}.

\bibitem[{Doogan and Buntine(2021)}]{doogan-buntine-2021-topic}
Caitlin Doogan and Wray Buntine. 2021.
\newblock \href {https://doi.org/10.18653/v1/2021.naacl-main.300} {Topic model
  or topic twaddle? re-evaluating semantic interpretability measures}.
\newblock In \emph{Proceedings of the 2021 Conference of the North American
  Chapter of the Association for Computational Linguistics: Human Language
  Technologies}, pages 3824--3848, Online. Association for Computational
  Linguistics.

\bibitem[{Hope et~al.()Hope, Hughes, Doshi-Velez, and
  Sudderth}]{hopeprediction}
Gabriel Hope, Michael~C Hughes, Finale Doshi-Velez, and Erik~B Sudderth.
\newblock Prediction-constrained hidden markov models for semi-supervised
  classification.

\bibitem[{Hoyle et~al.(2021)Hoyle, Goel, Hian-Cheong, Peskov, Boyd-Graber, and
  Resnik}]{hoyle2021automated}
Alexander Hoyle, Pranav Goel, Andrew Hian-Cheong, Denis Peskov, Jordan
  Boyd-Graber, and Philip Resnik. 2021.
\newblock Is automated topic model evaluation broken? the incoherence of
  coherence.
\newblock \emph{Advances in Neural Information Processing Systems}, 34.

\bibitem[{Hughes et~al.(2018)Hughes, Hope, Weiner, McCoy, Perlis, Sudderth, and
  Doshi-Velez}]{pmlr-v84-hughes18a}
Michael Hughes, Gabriel Hope, Leah Weiner, Thomas McCoy, Roy Perlis, Erik
  Sudderth, and Finale Doshi-Velez. 2018.
\newblock \href {https://proceedings.mlr.press/v84/hughes18a.html}
  {Semi-supervised prediction-constrained topic models}.
\newblock In \emph{Proceedings of the Twenty-First International Conference on
  Artificial Intelligence and Statistics}, volume~84 of \emph{Proceedings of
  Machine Learning Research}, pages 1067--1076. PMLR.

\bibitem[{Huh and Fienberg(2012)}]{huh2012discriminative}
Seungil Huh and Stephen~E Fienberg. 2012.
\newblock Discriminative topic modeling based on manifold learning.
\newblock \emph{ACM Transactions on Knowledge Discovery from Data (TKDD)},
  5(4):1--25.

\bibitem[{Iwata(2021)}]{iwata2021few}
Tomoharu Iwata. 2021.
\newblock Few-shot learning for topic modeling.
\newblock \emph{arXiv preprint arXiv:2104.09011}.

\bibitem[{Jordan et~al.(1999)Jordan, Ghahramani, Jaakkola, and
  Saul}]{jordan1999introduction}
Michael~I Jordan, Zoubin Ghahramani, Tommi~S Jaakkola, and Lawrence~K Saul.
  1999.
\newblock An introduction to variational methods for graphical models.
\newblock \emph{Machine learning}, 37(2):183--233.

\bibitem[{Kingma et~al.(2014)Kingma, Rezende, Mohamed, and
  Welling}]{kingma2014semisupervised}
Diederik~P. Kingma, Danilo~J. Rezende, Shakir Mohamed, and Max Welling. 2014.
\newblock \href {http://arxiv.org/abs/1406.5298} {Semi-supervised learning with
  deep generative models}.

\bibitem[{Kingma and Welling(2014)}]{kingma2014autoencoding}
Diederik~P Kingma and Max Welling. 2014.
\newblock \href {http://arxiv.org/abs/1312.6114} {Auto-encoding variational
  bayes}.

\bibitem[{Kingma et~al.(2015)Kingma, Salimans, and
  Welling}]{kingma2015variational}
Durk~P Kingma, Tim Salimans, and Max Welling. 2015.
\newblock Variational dropout and the local reparameterization trick.
\newblock \emph{Advances in neural information processing systems}, 28.

\bibitem[{Lacoste-Julien et~al.(2008)Lacoste-Julien, Sha, and
  Jordan}]{lacoste2008disclda}
Simon Lacoste-Julien, Fei Sha, and Michael Jordan. 2008.
\newblock Disclda: Discriminative learning for dimensionality reduction and
  classification.
\newblock \emph{Advances in neural information processing systems}, 21.

\bibitem[{Liu et~al.(2016)Liu, Tang, Dong, Yao, and Zhou}]{liu2016overview}
Lin Liu, Lin Tang, Wen Dong, Shaowen Yao, and Wei Zhou. 2016.
\newblock An overview of topic modeling and its current applications in
  bioinformatics.
\newblock \emph{SpringerPlus}, 5(1):1--22.

\bibitem[{Mandt et~al.(2016)Mandt, McInerney, Abrol, Ranganath, and
  Blei}]{mandt2016variational}
Stephan Mandt, James McInerney, Farhan Abrol, Rajesh Ranganath, and David Blei.
  2016.
\newblock Variational tempering.
\newblock In \emph{Artificial intelligence and statistics}, pages 704--712.
  PMLR.

\bibitem[{Mao et~al.(2012)Mao, Ming, Chua, Li, Yan, and
  Li}]{DBLP:conf/emnlp/MaoMCLYL12}
Xianling Mao, Zhaoyan Ming, Tat-Seng Chua, Si~Li, Hongfei Yan, and Xiaoming Li.
  2012.
\newblock \href {http://www.aclweb.org/anthology/D12-1073} {Sshlda: A
  semi-supervised hierarchical topic model}.
\newblock In \emph{EMNLP-CoNLL}, pages 800--809.

\bibitem[{Miao et~al.(2016)Miao, Yu, and Blunsom}]{miao2016neural}
Yishu Miao, Lei Yu, and Phil Blunsom. 2016.
\newblock \href {http://arxiv.org/abs/1511.06038} {Neural variational inference
  for text processing}.

\bibitem[{Nan et~al.(2019)Nan, Ding, Nallapati, and Xiang}]{nan2019topic}
Feng Nan, Ran Ding, Ramesh Nallapati, and Bing Xiang. 2019.
\newblock \href {http://arxiv.org/abs/1907.12374} {Topic modeling with
  wasserstein autoencoders}.

\bibitem[{Petinot et~al.(2011)Petinot, McKeown, and
  Thadani}]{petinot-etal-2011-hierarchical}
Yves Petinot, Kathleen McKeown, and Kapil Thadani. 2011.
\newblock \href {https://aclanthology.org/P11-2118} {A hierarchical model of
  web summaries}.
\newblock In \emph{Proceedings of the 49th Annual Meeting of the Association
  for Computational Linguistics: Human Language Technologies}, pages 670--675,
  Portland, Oregon, USA. Association for Computational Linguistics.

\bibitem[{Ramage et~al.(2009)Ramage, Hall, Nallapati, and
  Manning}]{10.5555/1699510.1699543}
Daniel Ramage, David Hall, Ramesh Nallapati, and Christopher~D. Manning. 2009.
\newblock Labeled lda: A supervised topic model for credit attribution in
  multi-labeled corpora.
\newblock In \emph{Proceedings of the 2009 Conference on Empirical Methods in
  Natural Language Processing: Volume 1 - Volume 1}, EMNLP '09, page 248–256,
  USA. Association for Computational Linguistics.

\bibitem[{Ren et~al.(2020)Ren, Kunes, and Doshi-Velez}]{ren2020prediction}
Jason Ren, Russell Kunes, and Finale Doshi-Velez. 2020.
\newblock Prediction focused topic models via feature selection.
\newblock In \emph{International Conference on Artificial Intelligence and
  Statistics}, pages 4420--4429. PMLR.

\bibitem[{Rezende et~al.(2014)Rezende, Mohamed, and
  Wierstra}]{rezende2014stochastic}
Danilo~Jimenez Rezende, Shakir Mohamed, and Daan Wierstra. 2014.
\newblock Stochastic backpropagation and approximate inference in deep
  generative models.
\newblock In \emph{International conference on machine learning}, pages
  1278--1286. PMLR.

\bibitem[{Rosen-Zvi et~al.(2012)Rosen-Zvi, Griffiths, Steyvers, and
  Smyth}]{rosen2012author}
Michal Rosen-Zvi, Thomas Griffiths, Mark Steyvers, and Padhraic Smyth. 2012.
\newblock The author-topic model for authors and documents.
\newblock \emph{arXiv preprint arXiv:1207.4169}.

\bibitem[{Schneider et~al.(2017)Schneider, Fechner, Landrum, and
  Stiefl}]{doi:10.1021/acs.jcim.7b00249}
Nadine Schneider, Nikolas Fechner, Gregory~A. Landrum, and Nikolaus Stiefl.
  2017.
\newblock \href {https://doi.org/10.1021/acs.jcim.7b00249} {Chemical topic
  modeling: Exploring molecular data sets using a common text-mining approach}.
\newblock \emph{Journal of Chemical Information and Modeling},
  57(8):1816--1831.
\newblock PMID: 28715190.

\bibitem[{Sharma et~al.(2021)Sharma, Zeng, Narayanan, Parbhoo, and
  Doshi-Velez}]{sharma2021learning}
Abhishek Sharma, Catherine Zeng, Sanjana Narayanan, Sonali Parbhoo, and Finale
  Doshi-Velez. 2021.
\newblock On learning prediction-focused mixtures.
\newblock \emph{arXiv preprint arXiv:2110.13221}.

\bibitem[{Srivastava and Sutton(2017)}]{srivastava2017autoencoding}
Akash Srivastava and Charles Sutton. 2017.
\newblock \href {http://arxiv.org/abs/1703.01488} {Autoencoding variational
  inference for topic models}.

\bibitem[{Wang and Yang(2020)}]{wang2020neural}
Xinyi Wang and Yi~Yang. 2020.
\newblock Neural topic model with attention for supervised learning.
\newblock In \emph{International Conference on Artificial Intelligence and
  Statistics}, pages 1147--1156. PMLR.

\bibitem[{Wenzel et~al.(2020)Wenzel, Roth, Veeling, Swiatkowski, Tran, Mandt,
  Snoek, Salimans, Jenatton, and Nowozin}]{wenzel2020good}
Florian Wenzel, Kevin Roth, Bastiaan Veeling, Jakub Swiatkowski, Linh Tran,
  Stephan Mandt, Jasper Snoek, Tim Salimans, Rodolphe Jenatton, and Sebastian
  Nowozin. 2020.
\newblock How good is the bayes posterior in deep neural networks really?
\newblock In \emph{International Conference on Machine Learning}, pages
  10248--10259. PMLR.

\bibitem[{Yan et~al.(2013)Yan, Guo, Lan, and Cheng}]{yan2013biterm}
Xiaohui Yan, Jiafeng Guo, Yanyan Lan, and Xueqi Cheng. 2013.
\newblock A biterm topic model for short texts.
\newblock In \emph{Proceedings of the 22nd international conference on World
  Wide Web}, pages 1445--1456.

\bibitem[{Zweig and Weinshall(2013)}]{zweig2013hierarchical}
Alon Zweig and Daphna Weinshall. 2013.
\newblock Hierarchical regularization cascade for joint learning.
\newblock In \emph{International Conference on Machine Learning}, pages 37--45.
  PMLR.

\end{thebibliography}
\bibliographystyle{acl_natbib}

\appendix
\section{Appendix}
\subsection{Optimization Procedure}

During optimization, there are three components of LI-NTM that are being trained: the encoder neural network, $\beta$ (the word distributions per label and topic), and the classifier neural network. We found randomly initializing all three trainable components and training them together lead to undesirable local minima (both perplexity and classification accuracy were undesirable). Instead, we consistently achieved our best results by first training the classifier normally on the task before training all three components together. All experimental results shown used this optimization procedure.

\subsection{Embedded Topic Model (ETM)} \label{sssec:etm}
Please find the generative process for ETM below \cite{dieng2019topic}. Note that ETM has two latent dimensions. There is the $L$-dimensional embedding space which the vocabulary is embedded into and each document is represented by $K$ latent topics. Furthermore, note that in ETM, each topic is represented by a vector $\alpha_k \in \mathbb{R}^L$ which is the embedded representation of the topic in embedding space. Furthermore, ETM defines an embedding matrix $\rho$ with dimension $L \times K$ where the column $\rho_v$ is the embedding of word $v$.
\begin{enumerate}
  \item Draw topic proportions $\theta_d \sim  \mathcal{LN}(0, I)$
  \item For each word $n$ in document:
  \begin{enumerate}
     \item Draw topic assignment $z_{dn} \sim $ Cat($\theta_d$)
     \item Draw word $w_{dn} \sim $ softmax($\rho^T \alpha_{z_{dn}}$)
  \end{enumerate}
\end{enumerate}

\subsection{Visualization of Topics}
See \autoref{fig:topwords}
\renewcommand{\thefigure}{A\arabic{figure}}
\setcounter{figure}{0}
\begin{figure*}[h]
\begin{center}
\includegraphics[scale=0.5]{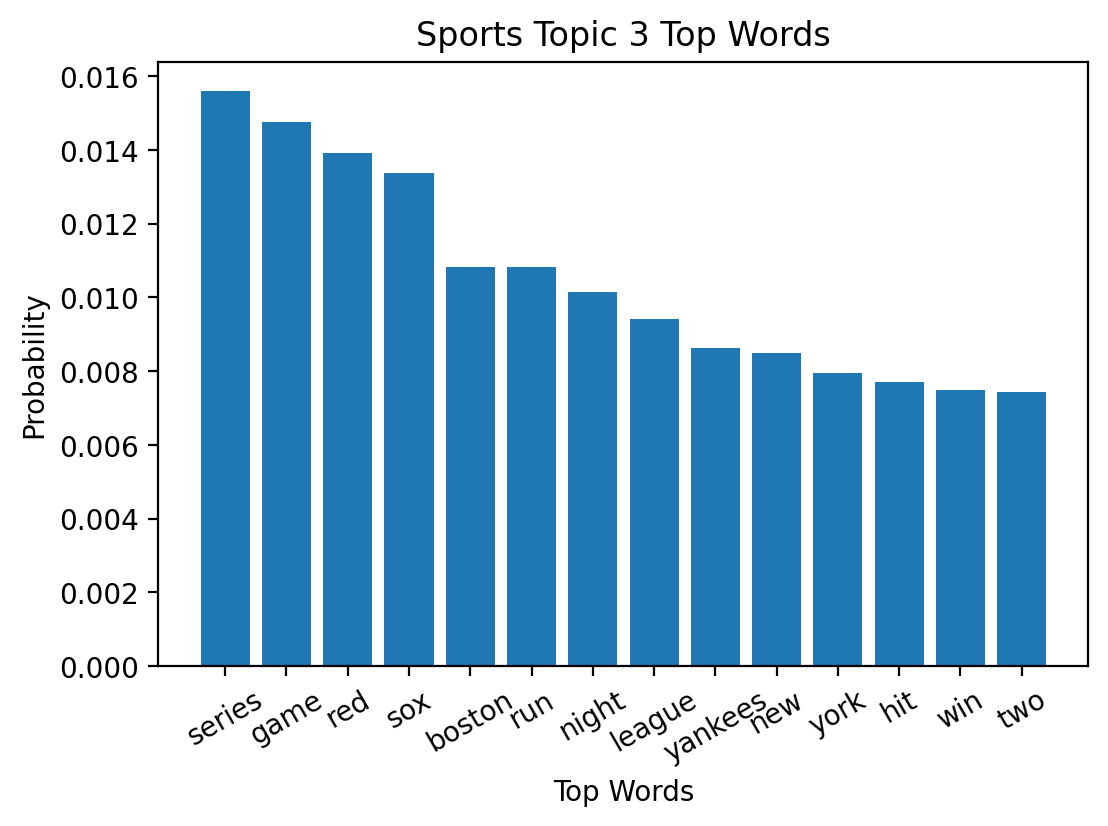}
\includegraphics[scale=0.5]{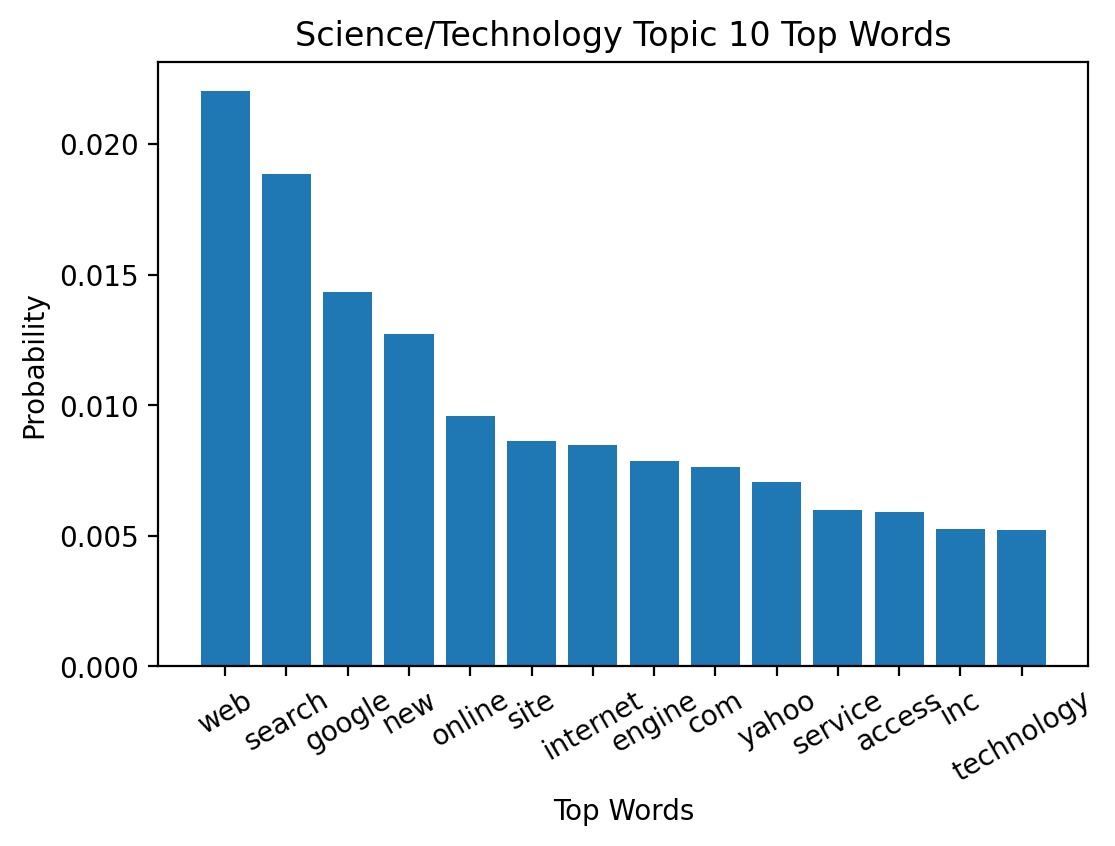}
\includegraphics[scale=0.5]{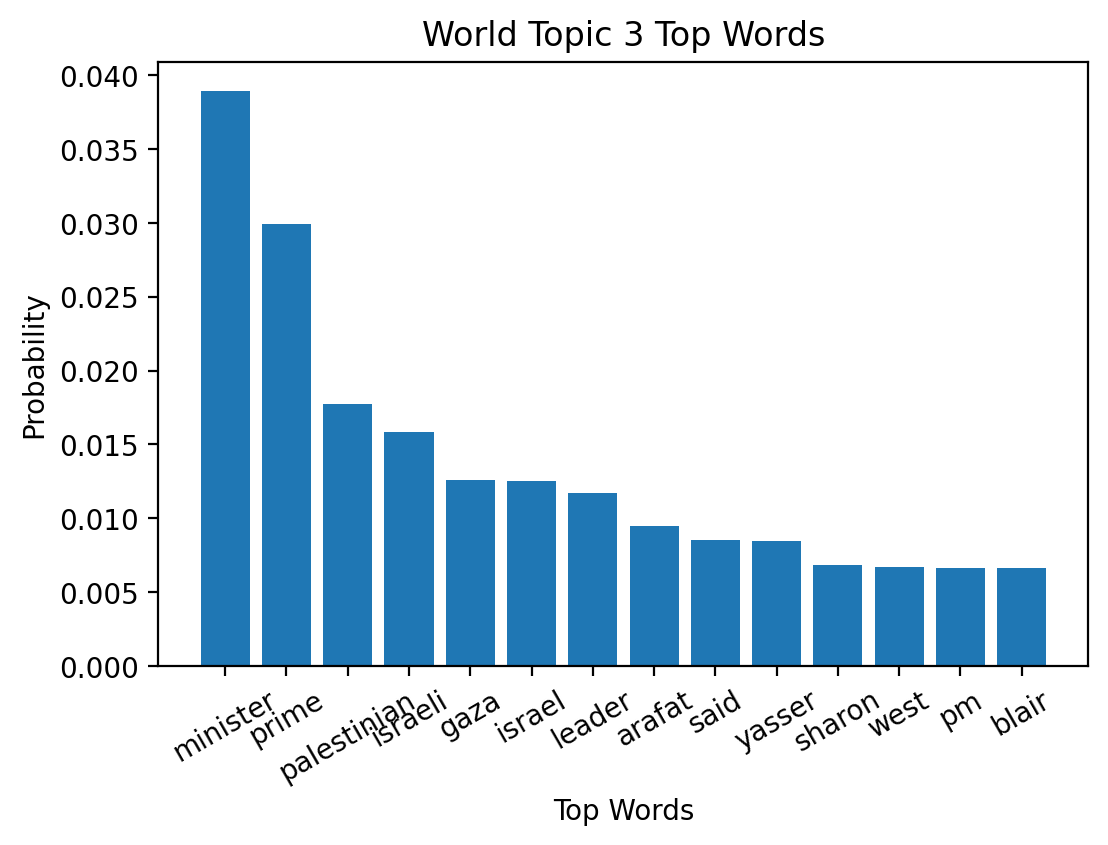}
\end{center}
   \caption{The probabilities of the top words from 5 selected topics from  LINT-m. Note that LINT-m has the nice property that topics are naturally sorted by label unlike ETM. The first topic, with words like "series", "yankees", "red", "sox", corresponds to baseball. Note that perplexity will still be high even if this topic is correctly given a high proportion in a baseball-themed news article since there are many potential baseball teams and baseball terminology that the article could be referencing. The second topic corresponds to search engines, and the third corresponds to the Israeli–Palestinian conflict.}
\label{fig:topwords}
\end{figure*}

\end{document}